\documentclass[]{elsarticle}
\usepackage{amsmath,amssymb,bm,color}
\usepackage{graphicx}
\usepackage{graphicx}
\usepackage{hyperref}
\usepackage{color}
\journal{Annals of Physics (published)}

\def\sumint{\int \!\!\!\!\!\!\!\!\! \sum }
\def\sint{\int \!\!\!\!\!\! \sum }

\begin{document}

\title{Nonuniversal equation of state for Rabi-coupled bosonic gases: a droplet
phase}
\author{Emerson Chiquillo}
\ead{emerson.chiquillo@unesp.br}

\address{Universidade Estadual Paulista (UNESP), Instituto de F\'isica
Te\'orica, S\~ao Paulo}

\begin{abstract}

Through an effective quantum field theory including zero temperature Gaussian 
fluctuations we derive analytical and explicit expressions for the equation of 
state of three-dimensional ultracold Rabi-coupled two-component bosonic gases 
with nonuniversal corrections to the interactions.
At mean-field level the system presents two ground-states, one symmetric and one 
non-symmetric or unbalanced.
For the symmetric ground state, in the regime where inter-species interactions 
are weakly attractive and subtly higher than repulsive intra-species, the 
instability by collapse is avoided by the contribution arising from Gaussian 
fluctuations, driving thus to formation of a liquidlike phase or droplet phase.
This self-bound state is crucially affected by the dependence on the nonuniversal
corrections to the interactions, which acts controlling the droplet stability.
By tuning the ratio between the inter-species scattering length and the 
intra-species scattering lengths or the nonuniversal contribution to the 
interactions we address and establish conditions under which the formation and 
stability of self-bound Rabi-coupled droplets with nonuniversal corrections to 
the interactions is favorable.

\end{abstract}

\maketitle

\section{Introduction}

The coupling and handling among atomic hyperfine states by means of laser beams
have become in a powerful tool leading to the possibility of inducing artificial 
transitions among such states \cite{Artificial,S1,S2}.
An extensive experimental, theoretical and numerical research has been addressed 
to understand the properties of these synthetic non-Abelian gauge fields in 
neutral bosonic mixtures of ultracold gases \cite{S3,S4,S5,S6,E1,S7,S8,E2,E3}.
It is remarkable to mention that most of theoretical developments have been 
carried out within the mean-field approximation by employing coupled 
Gross-Pitaevskii equations.
As a result of these breakthroughs there is currently a wide field of research in
spin-orbit- and Rabi-coupled ultracold bosonic atoms.
A few years after arising of these artificial systems, and in the context of 
Gaussian fluctuations in Bose-Bose mixtures a new matter phase was proposed 
\cite{Petrov1,Petrov2}.
This intriguing liquidlike state or droplet state lies in the highly nontrivial 
and subtle mechanical stabilization of paradigmatic mean-field collapse
threshold in attractive condensed Bose-Bose mixtures through the repulsive 
Gaussian correction coming from zero-point motion of Bogoliubov excitations.
Such a balance has revealed thus the crucial role played by fluctuations.
This droplet phase has allowed the birth and fast development of a new field of 
research \cite{Rabi,Drop-1d,Swirling,Bose-Bose,1D-3D-mixtures,Review,Rabi-E,
Adhikari2, Collective-excit, Pairing1,Pairing2,Phonon1,Normal-superfluid,
Boudjem, Bose-Bose2,Evaporation,Phonon2, Spinor1,Spinor2, Criticality,DNUC}.
Taking into account also the study of vortices, for example, in 
Ref. \cite{Vortices2d} it was shown that due to the effect of the logarithmic
factor in two-dimensional droplets, the stable vortex droplets manifest a 
flat-top shape with embedded vorticity between 1 and 5.
The spontaneous symmetry breaking in one-dimensional droplets loaded in a
symmetric double-core cigar-shaped potential is also considered in 
Ref. \cite{Symm-break-1d}. The results show droplets feature flat-top profiles
for large values of total norm, and bell (sech)-shaped density profiles for 
smaller values of total norm.

It is noteworthy that though the fluctuations are usually small, these can be
experimentally tuning by means of Feshbach resonances.
Thus experimental evidence has been obtained regarding outstanding influence of 
this small corrections.
Experimental droplets has been obtained in $^{39}$K-$^{39}$K mixtures
\cite{Exper-droplet1,Exper-droplet2,Exper-droplet3,Exper-droplet4} and 
$^{41}$K-$^{87}$Rb mixtures \cite{Exper-droplet5}.
Also has been studied the dynamical formation of self-bound droplets in 
attractive mixture of $^{39}$K atoms \cite{Exper-droplet6}.
More recently the observation of a Lee-Huang-Yang (LHY) fluid in a $^{39}$K spin 
mixture confined in a spherical trap potential has been achieved
\cite{Exper-droplet7}.
Droplets also have been obtained in a mixture of $^{23}$Na and $^{87}$Rb with
tunable attractive inter-species interactions \cite{Exper-droplet8}.
Shortly after the theoretical proposal of Bose-Bose droplets, this idea was
extended both at experimental and theoretical level to the context of 
one-component Bose gases with magnetic atoms,
\cite{trapped-dip-LHY-1,trapped-dip-LHY-2,d-LHY-1,d-LHY-2,Montecarlo-dip1,
d-LHY-3,Rosensweig,Self-Drop-Dy,Trap-Drop-Dy-1,Trap-Drop-Dy-2,Erbium}.
More recently a numerical study of stable 
three-dimensional anisotropic dipolar droplets with embedded vorticity and 
composite states with the vortex-antivortex-vortex structure is performed in Ref.
\cite{Vortices3d-dip}.
However, fitting experimental results with theoretical predictions some 
discrepancies have been found for both Bose-Bose droplets and droplets in 
bosonic dipolar gases.
In fact, in three-dimensional Bose-Bose gases it was observed that a minimum 
number of atoms is required to achieve a stable droplet formation
\cite{Exper-droplet1}, and this value is less than predicted in Ref.
\cite{Petrov1}.
Some differences have also been found between the theoretical model and Monte
Carlo results in Bose-Bose droplets
\cite{Numeric-drop,Montecarlo-Bose-Bose1,Montecarlo-Bose-Bose2}.
In an attempt to achieve a better description of this phase a phenomenological
beyond Lee-Huang-Yang (LHY) framework has been developed in \cite{Beyond LHY}, 
for Bose-Bose droplets in three and one dimension.
Therefore, given the discrepancies with both experimental and numerical results,
a better description of this phenomenon remains as an open issue.
From a theoretical framework, it is worth noting that both Bose-Bose droplets and
bosonic dipolar droplets remain weakly interacting, allowing thus for a
theoretical perturbative treatment.
In such a scenario for two-component bosonic droplets a key ingredient is to
consider the weakly interactions approximated by a local contact interaction.
In this sense, only the s-wave scattering length characterizes the two-body 
interatomic interactions, and we have an universal regime \cite{Andersen}. 
On the other hand, if the physical quantities depend on properties other than 
the s-wave scattering length we have a nonuniversal regime.
In addition and due to the sensitive nature of the fluctuations themselves the 
effect of some other parameter on these remains as an open issue.

Therefore taking into account the above mentioned, a connection between 
artificial spinor condensates and droplets is an interesting not widely explored
research field.
In this sense a first attempt to connect these two fields of research has been
carried out in Refs. \cite{Rabi,Rabi-E}. In these works  the coherent coupling 
between two atomic internal states provided by Rabi coupling and droplets in 
two-component Bose gases of interacting alkali-metal atoms was investigated. 
More recently, experimental measures of fluctuations with vanishing mean-field
energy in an asymmetric Rabi-coupled Bose-Einstein condensate with atoms of
$^{39}$K in a waveguide have been carried out in Ref. \cite{Exp-Rabi}.
So, for a better understanding of fluctuations behavior on Rabi-coupled 
Bose-Bose mixtures and motivated by the enhanced role of these, we consider an
extra ingredient.
We propose a step beyond by researching the effect of the nonuniversal 
corrections to the interactions \cite{Nonuniversal1, Nonuniversal2}
on Rabi-coupled bosonic droplets. 
Recently, Bose-Bose gases with such a kind of corrections to the interactions in
three, two, and one dimension both at zero and finite temperature have been 
considered in Ref. \cite{DNUC}.
It has been shown as the nontrivial dependence on the nonuniversal corrections 
to the interactions crucially affects the self-bound droplets stability. 
Hence we address theoretically on the zero temperature nonuniversal equation of
state for three-dimensional Rabi-coupled bosonic gases and we focus on the 
formation and stability of self-bound Rabi-coupled Bose-Bose droplets
with nonuniversal corrections to the interactions.

The rest of the paper is organized as follows.
In Section \ref{Rabi-coupled bosons} we introduce an effective-field theory 
in the path-integral formalism to describe interacting Rabi-coupled bosons with 
nonuniversal corrections to the interactions. We derive the mean-field 
grand-canonical potential, and we establish some conditions under which it is 
possible to obtain analytical expressions for the ground state.
In fact, we present two different kind of ground states one symmetric and one
non-symmetric or unbalanced.
We calculate the leading contribution to the equation of state for 
the symmetric ground-state at level of the Gaussian fluctuations at zero 
temperature in section \ref{Gaussian fluctuations of the symmetric ground-state}.
In order to obtain closed analytical expressions useful simplifications take
place when, in the spectrum, we make use of the physically reasonable 
approximation of weak Rabi frequency, $\omega_R \ll \mu/\hbar$ 
\cite{Magnetic-solit-Rabi}. 
We perform dimensional regularization to solve the divergent zero-point integrals
due to both gapless and gapped elementary excitations.
Here we analyze the conditions for the existence of a stable formation of 
droplets with attractive inter-species scattering length and repulsive 
intra-species scattering length. 
In such a case, we obtain an analytical expression for the energy density of 
nonuniversal Rabi-coupled bosonic droplets with a nontrivial dependence of 
scattering lengths, nonuniversal effects of the interactions, and the Rabi 
coupling. We also include some phase diagrams relating some of the parameters 
above mentioned.
In section \ref{Summary and outlook} we draw a final discussion and we present
some future perspectives of our results.

\section{An effective-field theory for Nonuniversal Rabi-coupled bosons}
\label{Rabi-coupled bosons}

In this section we introduce some elements of interest in the
path-integral formalism. 
This functional formalism is equivalent to the creation and annihilation 
operators procedure \cite{Le Bellac,Stoof}. Fluctuations in three-dimensional 
($3d$) one-component bosonic gases with hard-sphere interactions were originally 
considered in \cite{LHY-1,LHY-2}. In such a work the LHY-corrections were
obtained using the formalism of creation and annihilation operators.
In the context of fluctuations in two-component bosonic gases, the same operators
procedure it was considered in Refs. \cite{Petrov1,Petrov2,Larsen}.
On the other hand, same results were obtained into the path-integral formalism in
Ref. \cite{1D-3D-mixtures}.
So, we consider $3d$ equal-mass two-component Rabi-coupled ultracold bosonic 
gases with nonuniversal corrections to the interactions and chemical potential 
$\mu$, in the path-integral formalism. 
The action in a $3d$ box of volume $V$ can be read as \cite{Rabi,DNUC}
\begin{eqnarray}
S[\Psi,\Psi^*] &=&  \sumint
\big[\psi_\alpha^* (\textbf{r},\tau) \big(\hbar\frac{\partial}{\partial\tau}
- \frac{\hbar^2}{2m}\nabla^2  -\mu\big)\psi_\alpha(\textbf{r},\tau)
\nonumber \\
&+& \frac{1}{2}\sum_{\sigma}g^{(0)}_{\alpha\sigma}
|\psi_\alpha(\textbf{r},\tau)|^2|\psi_\sigma(\textbf{r},\tau)|^2
\nonumber \\
&-& \frac{1}{2}\sum_{\sigma}g^{(2)}_{\alpha\sigma}
|\psi_\alpha(\textbf{r},\tau)|^2 \nabla^2 |\psi_\sigma(\textbf{r},\tau)|^2
\nonumber \\
&-&  \hbar\omega_R(\psi_1^*\psi_2 + \psi_2^*\psi_1)\Big],
\label{S}
\end{eqnarray}
where we have used the shorthand notation 
$\,\sint \equiv\int_0^{\hbar \beta} d\tau \int_{V}  d^3 r
\sum_{\alpha}$, $\beta = 1/k_B T$, $k_B$ is the Boltzmann constant, and the
hyperfine states are labeled as $\alpha,\sigma=1,2$.
Each component is described by a complex bosonic field $\psi_\alpha$, and these
fields are considered at position $\mathbf{r}$ and imaginary time $\tau$.
The superscripts $(0,2)$ are related to the zero-range approximation to the
interactions and the nonuniversal improvements to the interactions, respectively.
Dealing with ultracold and dilute bosons, the most common scheme to taking into
account the interactions is to consider an approximated zero-range potential
\cite{Pit-Strin,Peth-Smith}. 
Hence, the strength of the interactions is expressed through the intra- and
interspecies coupling constants 
$g^{(0)}_{\alpha\alpha}=4\pi\hbar^2a_{\alpha\alpha}/m$ and
$g^{(0)}_{\alpha\sigma}=4\pi\hbar^2a_{\alpha\sigma}/m$, respectively.
These in turn are related to the $s$-wave scattering lengths
$a_{\alpha\alpha}$ and $a_{\alpha\sigma}$, respectively, such that
$a_{\alpha\alpha},a_{\alpha\sigma}>0$ represent repulsion, and
$a_{\alpha\alpha},a_{\alpha\sigma}<0$ attraction.
In order to research the role played by the effects beyond the zero-range
approximation to the interactions, we include the nonuniversal corrections to the
two-body interatomic interaction potential in the third line of Eq. (\ref{S}).
Thus we have the coupling constants for intra-species and inter-species with 
nonuniversal corrections to the interactions
$g^{(2)}_{\alpha\alpha}$ and $g^{(2)}_{\alpha\sigma}$, respectively. 
It is remarkable mentioning that from scattering theory there are two different
coupling constants considering the s-wave effective range of interaction $r$,
such that \cite{Scattering1, Scattering2, On-shell}
\begin{eqnarray}
g^{(2)}_{\alpha\alpha} = \frac{2\pi\hbar^2}{m}a^2_{\alpha\alpha}r_{\alpha\alpha} 
\quad \quad \quad \quad a_{\alpha\alpha}\gg r_{\alpha\alpha},
\label{Coupl1}
\end{eqnarray}
\begin{eqnarray}
g^{(2)}_{\alpha\sigma} = \frac{2\pi\hbar^2}{m}a^2_{\alpha\sigma}r_{\alpha\sigma}
\quad \quad \quad \quad a_{\alpha\sigma}\gg r_{\alpha\sigma},
\label{Coupl2}
\end{eqnarray}
and, \cite{Adhikari}
\begin{eqnarray}
g^{(2)}_{\alpha\alpha} = -\frac{2\pi\hbar^2}{3m}r^3_{\alpha\alpha}
\quad \quad \quad \quad a_{\alpha\alpha}\ll r_{\alpha\alpha},
\label{fr-1}
\end{eqnarray}
\begin{eqnarray}
g^{(2)}_{\alpha\sigma} = -\frac{2\pi\hbar^2}{3m}r^3_{\alpha\sigma}
\quad \quad \quad \quad a_{\alpha\sigma} \ll r_{\alpha\sigma}.
\label{fr-2}
\end{eqnarray}
These last with a good numerical description of low-energy scattering.
In absence of the Rabi coupling, a better description of the droplet phase is
achieved by considering the couplings (\ref{fr-1}), and (\ref{fr-2}), 
see Ref. \cite{DNUC}. From now on we will take into account these couplings. 
Transitions between the two states are induced by an external coherent Rabi
coupling of frequency $\omega_R>0$.
Due to the Rabi mixing between states, only the total number of particles is 
conserved \cite{Son}. Thus it is assumed that the two components are in a state 
with the same chemical potential $\mu$ \cite{Son,Equal-chem-pot, Rabi1}.

In order to obtain the ground state of the system, we calculate the grand 
potential $\Omega = -\beta^{-1} \ln\mathcal{Z}$. 
In the path-integral formalism the grand canonical partition function
$\mathcal{Z}$ at temperature $T$ is written as
\begin{eqnarray}
\mathcal{Z} = \int \mathcal{D}[\Psi,\Psi^*]\text{exp}(-S[\Psi,\Psi^*]/\hbar).
\end{eqnarray}
At zero temperature we assume the bosons condensate into the zero-momentum 
states. In other words, we consider the superfluid phase, where a $U(1)$ gauge 
symmetry of each component is spontaneously broken. So in order to performing a 
perturbative expansion we set $\psi_\alpha (\textbf{r},\tau) = \phi_{\alpha}
+ \eta_\alpha (\textbf{r},\tau)$ \cite{Rabi}. 
Where $\phi_{\alpha} \equiv \langle \psi_\alpha (\textbf{r},\tau) \rangle$ 
\cite{Stoof}, corresponds to the condensate wave-function.
The fluctuations around $\psi_{\alpha}$ are given by
$\eta_\alpha (\textbf{r},\tau)$ and these are orthogonal to the condensate of the
same species \cite{Stoof, Modes}.
Here we have $n_{\alpha}=|\phi_{\alpha}|^2=N_\alpha/L^d$, as the density of 
particles in the mean-field approximation or macroscopic density of the
Bose-Einstein condensate.
After introducing $\psi_\alpha (\textbf{r},\tau) = \phi_{\alpha}
+ \eta_\alpha (\textbf{r},\tau)$ into the action (\ref{S}) we expand the action
up to the second order (Gaussian) in
$\eta_\alpha (\textbf{r},\tau)$ and $\eta^*_\alpha (\textbf{r},\tau)$, and we 
arrive at the split grand potential $\Omega= \Omega_0 + \Omega_{\text{f}}$. Where
$\Omega_0$ is the mean-field contribution, while $\Omega_{\text{f}}$ takes into 
account the Gaussian zero-temperature fluctuations.
Since $\phi_{\alpha}$ describes the condensate, the linear terms in 
the fluctuations vanish such that $\phi_{\alpha}$ really minimizes 
the action \cite{Stoof}.
Thus the explicit form of the mean-field grand-potential could be obtained by 
minimizing the mean-field grand potential $\Omega_0$ with respect to $\phi$
(saddle-point approximation), as will be seen later.

\subsection{Mean-field ground states}
At first instance we carry out the analysis of the mean-field contribution, where 
$\psi_\alpha = \phi_{\alpha}$, from which we get the following grand-potential
\begin{eqnarray}
\frac{\Omega_0}{V} = \sum_{\alpha=1,2} \big(-\mu \phi^2_\alpha
+ \frac{1}{2}\sum_{\sigma=1,2}g^{(0)}_{\alpha\sigma}
\phi^2_\alpha \phi^2_\sigma\big)
- 2\hbar\omega_R \phi_1\phi_2,
\label{Grand-0}
\end{eqnarray}
which is not dependent on the nonuniversal effects of the interactions, given the
derivative of the nonuniversal contribution in Eq. (\ref{S}).
In the saddle-point approximation i.e., minimizing $\Omega_0$ with respect to
$\phi_{\alpha}$, we obtain
\begin{eqnarray}
\mu\phi_\alpha = \big(g^{(0)}_{\alpha\alpha}\phi^2_\alpha 
+ g^{(0)}_{\alpha\sigma}\phi^2_\sigma\big)\phi_\alpha
- \hbar\omega_R \phi_\sigma,
\label{minim}
\end{eqnarray}
with $\alpha,\sigma=1,2$ and $\alpha\neq\sigma$.
In addition  useful simplifications take place when the intra-species are equal, 
such that, $g^{(0)}_{11}=g^{(0)}_{22}=g^{(0)}= 4\pi \hbar^2a/m$. 
Therefore solving  Eq. (\ref{minim}) we get
\begin{eqnarray}
\big[\big(g^{(0)} - g^{(0)}_{12}\big) \phi_1 \phi_2 + \hbar\omega_R \big]
(\phi^2_1 - \phi^2_2)=0.
\label{Eq}
\end{eqnarray}

Since $|\phi_1|^2= n_1$, and $|\phi_2|^2= n_2$ are the densities of each internal
state, the total density is given as $n=n_1 + n_2$.
Thus two solutions of Eq. (\ref{Eq}) are obtained.
One is a symmetric phase where the two internal states are equally populated 
$n_1=n_2$. The other solution is a non-symmetric or unbalanced configuration
which exists only in presence of the Rabi coupling and it arises when the intra- 
and inter-spin interactions of the two species are not equal, leading thus to a 
non-zero spin density polarization $n_1 - n_2$. 
Now briefly we describe some general aspects related to these ground-states
\cite{Rabi}.

\subsubsection{Symmetric ground-state}
\label{section}
In the symmetric phase we consider $n_1=n_2=n/2$, and from Eq. (\ref{minim}) we 
get
\begin{eqnarray}
n = \frac{2\mu_R}{g^{(0)} + g^{(0)}_{12}},
\label{Chem-symmetric}
\end{eqnarray}
where $\mu_R = \mu + \hbar \omega_R$, and the corresponding mean-field 
grand-potential can be read as
\begin{eqnarray}
\frac{\Omega_0}{V} = -\frac{\mu_R^2}{g^{(0)} + g^{(0)}_{12}},
\label{Mean-field-grand-symm}
\end{eqnarray}
with the respective mean-field energy density $\mathcal{E}_{0}$ given by
\begin{eqnarray}
\mathcal{E}_{0}=
\frac{1}{4}
\big( g^{(0)} + g^{(0)}_{12} \big)n^2 - \hbar\omega_R n,
\label{Ener-dens-symm}
\end{eqnarray}
where we can see that the minimum value of the energy density to obtain the 
stable symmetric ground-state 
$\mathcal{E}_{0,\text{min}}= -\hbar^2\omega^2_R/(g^{(0)} + g^{(0)}_{12})$
is handled by the Rabi frequency.
In this symmetric ground-state the Hessian matrix indicates that the 
grand-potential stability at mean-field level in Eq. (\ref{Grand-0}) is achieved 
satisfying the condition
\begin{eqnarray}
\big(g^{(0)} + g^{(0)}_{12}\big) \big[\big(g^{(0)} - g^{(0)}_{12}\big)\mu
+ 2\hbar \omega_R g^{(0)} \big]>0.
\label{Symmetric}
\end{eqnarray}

\begin{figure}[t]
\begin{center}
\includegraphics[height=7.5cm, width=8.5cm, clip]{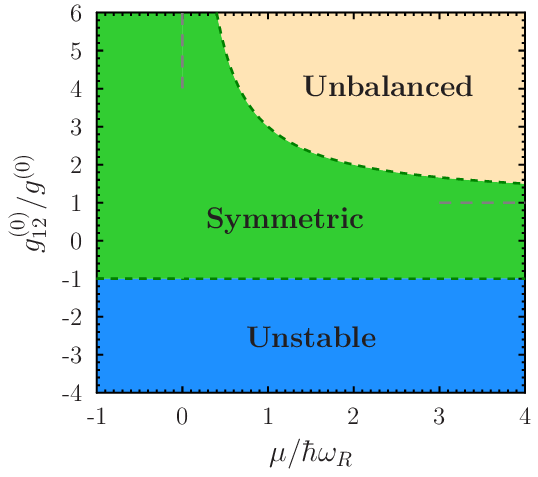}
\end{center}
\vskip -.6cm
\caption{Ground states mean-field phase diagram of interacting Rabi-coupled
bosons \cite{Rabi}.
Symmetric region is given by condition (\ref{Symmetric}), and the unbalanced or
non-symmetric region is obtained from condition (\ref{Non-symmetric}).
The ratio value $g^{(0)}_{12}/g^{(0)}=1$ represents the asymptotic behavior for 
large values of $\mu/\hbar \omega_R$ (horizontal dashed gray line), and
$\mu/\hbar \omega_R = 0$ represents the asymptotic behavior for large
$g^{(0)}_{12}/g^{(0)}$ ratio (vertical dashed gray line).
}
\label{F}
\end{figure}

In Fig. \ref{F} we show the existence condition for the symmetric ground-state
given by (\ref{Symmetric}).
In this phase, Eq. (\ref{Chem-symmetric}) establishes $\mu/\hbar\omega_R > -1$ in
order to have a physical density $n>0$.
Without Rabi coupling it is know that for a complete set of repulsive 
interactions, where $g^{(0)}>0$ and $g^{(0)}_{12}>0$, in order to ensure a stable
miscibility phase (overlapping of two components), and avoiding the phase
separation, it is necessary providing $0<g^{(0)}_{12}/g^{(0)}<1$ 
\cite{Pit-Strin,Peth-Smith}.
In the phase separation the repulsion between atoms belonging to the same species
is stronger that repulsion between species and as consequence there is not mixing between two species.
However, the inclusion of Rabi coupling gives rise to an effective attraction 
between the species which can drive the immiscible configuration into a miscible
state, and the stable ground-state remains.
On the other hand, at mean-field level a condensed Bose-Bose mixture in absence 
of Rabi coupling collapses when the interspecies attraction $g^{(0)}_{12}<0$ 
becomes stronger than intraspecies repulsion, $g^{(0)}>0$, providing
$[g^{(0)}_{12}]^2>[g^{(0)}]^2$ with $-1<g^{(0)}_{12}/g^{(0)} < 0$.  
In such a case this attractive mechanism is leading and there is not an effect
of the Rabi contribution. The mixture remains unstable for 
$g^{(0)}_{12}/g^{(0)}<-1$, blue region in Fig. \ref{F}.

\subsubsection{Unbalanced or Non-symmetric ground-state}

In the non-symmetric phase using the second solution from Eq. (\ref{Eq}) into
Eq. (\ref{minim}), the density of each state can be read as
\begin{eqnarray}
n_{1,2}= \frac{\mu}{2g^{(0)}}\bigg[ 1 \pm \sqrt {1- \Big[
\frac{2\hbar\omega_R g^{(0)}}{\big(g^{(0)} - g^{(0)}_{12}\big)\mu} \Big]^2 }
\,\bigg]
\label{Stab-Unbalanced}
\end{eqnarray}
such that $n=\mu/g^{(0)}$. 
The Rabi coupling presence destabilizes the symmetric ground-state allowing
thus the arising of this unbalanced phase.
The corresponding mean-field grand potential can be read as
\begin{eqnarray}
\frac{\Omega_0}{V} = -\frac{1}{2} \frac{\mu^2}{g^{(0)}}
+ \frac{(\hbar \omega_R)^2}{g^{(0)} - g^{(0)}_{12}},
\end{eqnarray}
with the respective mean-field energy density given by
\begin{eqnarray}
\mathcal{E}_{0}= \frac{1}{2}
g^{(0)} n^2 + \frac{(\hbar\omega_R)^2}{g^{(0)} - g^{(0)}_{12}},
\end{eqnarray}
where the Rabi coupling contribution acts as an shift of the mean-field 
energy density of a single Bose-Einstein condensate.
The coupling constants also impose the condition $g^{(0)}\neq g^{(0)}_{12}$
for the existence of the unbalanced ground-state.
From Eq. (\ref{Grand-0}) the stability of the non-symmetric ground-state is 
obtained providing
\begin{eqnarray}
\big(g^{(0)} - g^{(0)}_{12}\big) \Big[4g^{(0)}-
\frac{\big(g^{(0)} - g^{(0)}_{12}\big)^2\mu^2}{g^{(0)}(\hbar \omega_R)^2} 
\Big]>0,
\label{Non-symmetric}
\end{eqnarray}
as it is shown in the Fig. \ref{F}.

\section{Gaussian fluctuations of the symmetric ground-state}
\label{Gaussian fluctuations of the symmetric ground-state}

Hereafter, we focus on the symmetric configuration.
In addition to having the same intra-species couplings, we have also considered 
the simplified scenario where the intra-species nonuniversal effects to the 
interactions are equal, such that
$g_{11}^{(2)}=g_{22}^{(2)}=g^{(2)}= 2\pi\hbar^2 a^2r/m$.
The grand potential of the Gaussian fluctuations $\Omega_{\text{f}}$ is provided 
by \cite{Regularization}
\begin{eqnarray}
\Omega_{\text{f}} = \frac{1}{2\beta}\sum_{k>0}
\sum^{+\infty}_{\substack{n=-\infty}} \ln \det [\mathbb{G}^{-1}(k,\omega_n)],
\label{Grand1}
\end{eqnarray}
with the bosonic Matsubara's frequencies $\omega_n=2\pi n/ (\hbar \beta)$, and
the $4\times4$ inverse fluctuation propagator $\mathbb{G}^{-1}$ given as
\begin{equation}
\mathbb{G}^{-1} =
\begin{pmatrix}
\textbf{G}_{11}^{-1} & \textbf{G}_{12}^{-1} \\
\\
\textbf{G}_{12}^{-1} & \textbf{G}_{22}^{-1} \\
\end{pmatrix},
\label{invers-propagator}
\end{equation}
with the symmetric $2\times2$ matrices
\begin{equation}
\textbf{G}_{11}^{-1} =
\begin{pmatrix}
-\text{i}\hbar \omega_n +f_1(k)
& \frac{1}{2}(g^{(0)} + g^{(2)}k^2)\phi^2 \\
\\
\frac{1}{2}(g^{(0)} + g^{(2)}k^2)\phi^2 &
\text{i}\hbar \omega_n + f_1(k)
\\
\end{pmatrix}
,
\end{equation}
where $f_1(k) = \epsilon_k +
\frac{1}{2}(2g^{(0)} + g^{(0)}_{12} + g^{(2)}k^2)\phi^2 - \mu$, 
with the free-particle energy $\varepsilon_k=\hbar^2k^2/2m$,
$\textbf{G}_{22}^{-1} = \textbf{G}_{11}^{-1}(1 \leftrightarrow2),$
and
\begin{equation}
\textbf{G}_{12}^{-1} =
\begin{pmatrix}
\frac{1}{2}(g^{(0)}_{12} + g^{(2)}_{12}k^2)\phi^2 -\hbar \omega_R
& \frac{1}{2}(g^{(0)}_{12} + g^{(2)}_{12}k^2)\phi^2 \\
\\
\frac{1}{2}(g^{(0)}_{12} + g^{(2)}_{12}k^2)\phi^2 &
\frac{1}{2}(g^{(0)}_{12} + g^{(2)}_{12}k^2)\phi^2 -\hbar \omega_R
\\
\end{pmatrix}
.
\end{equation}
By solving the determinant of the inverse propagator we get
\begin{eqnarray}
\Omega_{\text{f}}= - \frac{1}{2\beta}\sum^{+\infty}_{\substack{k>0\\
n=-\infty}} \ln \big[(\hbar^2\omega_n ^2 + E_+^2)
(\hbar^2\omega_n ^2 + E^2_-)\big],
\label{Grand1}
\end{eqnarray}
with the Bogoliubov spectra given as 
\begin{eqnarray}
E_+ = \sqrt{(1+\mu_R \delta_+) \varepsilon^2_k + 2\mu_R \varepsilon_k},
\end{eqnarray}
and
\begin{eqnarray}
E_- = \sqrt{(1+\mu_R \delta_-) \varepsilon^2_k +
2A\varepsilon_k + B}
\end{eqnarray}
with $A= 2\hbar \omega_R + \mu_R(\lambda + \hbar\omega_R\alpha_-)$,
$B=4\hbar\omega_R(\hbar\omega_R + \lambda \mu_R)$,
\begin{eqnarray}
\delta_{\pm} = \frac{4m}{\hbar^2}
\Big(\frac{g^{(2)}\pm g^{(2)}_{12}}{g^{(0)}+g^{(0)}_{12}} \Big),
\hspace{2cm}
\lambda = \frac{g^{(0)} - g^{(0)}_{12}}{g^{(0)}+g^{(0)}_{12}}.
\label{constants}
\end{eqnarray}

The sum over the bosonic Matsubara's frequencies given by Eq. (\ref{Grand1}) 
has two contributions, one at zero-temperature  and the other at 
finite-temperature \cite{1D-3D-mixtures,Le Bellac,Regularization}.
We neglect the contribution at finite temperature considering only the
zero-temperature term which is given by 
\begin{eqnarray}
\Omega_\text{f}=\frac{1}{2}\sum_{k,\pm}E_{\pm}(k),
\end{eqnarray}
which in turn in the continuum limit $\sum_k\rightarrow V \int d^3 k/(2\pi)^3$, 
can be read as follows
\begin{eqnarray}
\frac{\Omega_\text{f}}{V} = \frac{\Omega^+_\text{f}}{V}
+ \frac{\Omega^-_\text{f}}{V}
\label{grand-pot}
\end{eqnarray}
with the ultraviolet divergent integrals 
\begin{eqnarray}
\frac{\Omega^+_\text{f}}{V}
=\int_0^\infty \frac{dk}{4\pi^2}\, k^{2} E_+ 
\label{grand-pot+},
\end{eqnarray}
and
\begin{eqnarray}
\frac{\Omega^-_\text{f}}{V}
=\int_0^\infty \frac{dk}{4\pi^2}\, k^{2} E_-.
\label{grand-pot-}
\end{eqnarray}
In addition the second integral has not a closed form.
Therefore in order to get some analytical results and to know some insights into
the underlying physics, we consider the limit of weak Rabi frequency, such that,
$\omega_R \ll \mu/\hbar$ \cite{Rabi,Magnetic-solit-Rabi}.
So by expanding up the Bogoliubov spectra to first order in the Rabi frequency, 
we get 
\begin{eqnarray}
E_+ = [\varepsilon^2_k(1+\delta_+\mu)+2\mu\varepsilon_k]^{1/2}
+ \frac{\varepsilon_k(2+\varepsilon_k \delta_+)\hbar\omega_R}
{2[\varepsilon^2_k(1+\delta_+\mu)+2\mu\varepsilon_k]^{1/2}}
+ \mathcal{O}\big(\big(\frac{\hbar\omega_R}{\mu}\big)^2\big),
\label{E+}
\end{eqnarray} 
and 
\begin{eqnarray}
E_- &=&  [\varepsilon^2_k(1+\delta_-\mu)+2\lambda\mu\varepsilon_k]^{1/2}
\nonumber \\
&+&  \frac{[\varepsilon^2_k \delta_- 
+ 2\varepsilon_k(2+\lambda+\mu\delta_-) + 4\lambda\mu]\hbar\omega_R}
{2[\varepsilon^2_k(1+\delta_-\mu)+2\lambda\mu\varepsilon_k]^{1/2}}
+ \mathcal{O}\Big(\Big(\frac{\hbar\omega_R}{\mu}\Big)^2\Big).
\label{E-}
\end{eqnarray}
The new ultraviolet divergent integrals are solved by means of dimensional 
regularization \cite{Rabi-E,Regularization}, and the resulting two branches of
the grand potential are
\begin{eqnarray}
\frac{\Omega^+_\text{f}}{V} = \frac{8}{15\pi^2} \Big(\frac{m}{\hbar^2}\Big)^{3/2}
\frac{\mu^{5/2}}{(1+\delta_+\mu)^2}
\Big[1+ \frac{1}{2}\Big(\frac{5+\delta_+\mu}{1+\delta_+\mu} 
\Big)\frac{\hbar\omega_R}{\mu}\Big],
\label{Omega+}
\end{eqnarray}
obtained from  Eq. (\ref{E+}), and 
\begin{eqnarray}
\frac{\Omega^-_\text{f}}{V}&=&\frac{8}{15\pi^2} \Big(\frac{m}{\hbar^2}\Big)^{3/2}
\frac{\lambda^{5/2}\mu^{5/2}}{(1+\delta_-\mu)^2}
\Big[ 1 + 
\frac{1}{4\lambda(1+\delta_-\mu)}  
\nonumber \\
&\times& 
\big[ 2\lambda(5+\delta_-\mu) + 5[1-(\delta_-\mu)^2] \big]
\frac{\hbar\omega_R}{\mu}
\Big].
\label{Omega-}
\end{eqnarray}
obtained from  Eq. (\ref{E-}).
We stress that the first integral in Eq. (\ref{grand-pot}) can be solved 
directly, for any value of Rabi frequency, by means of dimensional regularization
\cite{DNUC,Regularization,Hooft,dim-reg}, resulting 
\begin{eqnarray}
\frac{\Omega^+_\text{f}}{V} = \frac{8}{15\pi^2} \Big(\frac{m}{\hbar^2}\Big)^{3/2}
\frac{\mu_R^{5/2}}{(1+\delta_+\mu_R)^2},
\label{Omega+complet}
\end{eqnarray}
and the grand potential in Eq. (\ref{Omega+}) can be also obtained directly from
Eq. (\ref{Omega+complet}) in the limit of small Rabi coupling.
We stress that in absence of non-universal corrections Eq. (\ref{Omega+complet}) 
was considered in \cite{Rabi}.
In this reference the grand potential was used directly to obtain the third term 
in the energy given by Eq. 14, which is independent of the Rabi coupling. However the two
last contributions in such a energy are obtained in the approximation of weak
Rabi frequency (see also \cite{Rabi-E} for a one-dimensional version of this 
issue with the same consideration). Here, we calculate the energy from the two 
branches of the grand potential in the limit of small Rabi coupling given by 
Eqs. (\ref{Omega+}) and (\ref{Omega-}).
In this way we show that in absence of the non-universal contribution our results
for the energy do not completely match from those obtained in \cite{Rabi}.

Now we consider $E_B = \hbar^2/ma^2$ as energy unit, the Rabi-frequency
$\widetilde{\omega}_R E_B= \hbar \omega_R$, $\widetilde{n} = na^3$, the energy 
density $ \mathcal{E}=E/V$, the ratios $\epsilon = a_{12}/a$, 
$\gamma = r/a$, and $ \gamma_{12} = r_{12}/a$. This leads to the total scaled
energy density $\widetilde{\mathcal{E}} = \mathcal{E}/(E_B/a^3)$ for nonuniversal
Rabi-coupled interacting bosonic gases in the small Rabi frequency regime 
\begin{eqnarray}
\widetilde{\mathcal{E}} &=& \pi(1+\epsilon)\widetilde{n}^2
- \widetilde{\omega}_R\widetilde{n}
\nonumber \\
&+&\frac{32\sqrt{2\pi}}{15}
\frac{[(1+\epsilon)\widetilde{n}]^{5/2}} {(1-h^+_\text{fr}\widetilde{n})^2}
\bigg[1+\frac{\widetilde{\omega}_R}{4\pi(1+\epsilon)\widetilde{n}}
\bigg[\frac{5-h^+_\text{fr}\widetilde{n}}{1-h^+_\text{fr}\widetilde{n}}\bigg]
\bigg]
\nonumber \\
&+&
\frac{32\sqrt{2\pi}}{15} \frac{[(1-\epsilon)\widetilde{n}]^{5/2} }
{(1+h^-_\text{fr}\widetilde{n})^2} \bigg\{1+ \frac{\widetilde{\omega}_R}
{8\pi(1-\epsilon) \widetilde{n}} \frac{1} {(1+h^-_\text{fr}\widetilde{n})}
\nonumber \\
&\times& \Big[2\Big(\frac{1-\epsilon}{1+\epsilon}\Big)(5+h^-_\text{fr}
\widetilde{n})+5[1-(h^-_\text{fr}\widetilde{n})^2]\Big] \bigg\},
\label{TotalE}
\end{eqnarray}
where $h^\pm_\text{fr} = 4\pi(\gamma^3_{12} \pm \gamma^3)/3$. We have included 
the mean-field energy density given by Eq. (\ref{Ener-dens-symm}) in the first 
line.
The second line is obtained from Eq. (\ref{Omega+}), and the last two from 
Eq. (\ref{Omega-}).
By considering the couplings in Eqs. (\ref{Coupl1}) and (\ref{Coupl2}) it is 
necessary taking into account the change of variable
$\gamma^3_{12} \pm \gamma^3 \rightarrow \mp 3 (\gamma \pm \epsilon^2 
\gamma_{12})$. 

\subsection{Nonuniversal droplet phase}
The energy density in Eq. (\ref{TotalE}) is well-defined for $-1<\epsilon\leq 1$,
in correspondence with the prediction set by the mean-field analysis, as it was
discussed at end of section \ref{section}.
In absence of the Rabi coupling, the miscible phase, is achieved for 
$0<\epsilon<1$ \cite{Pit-Strin,Peth-Smith}, at mean-field level.
However the Rabi frequency allows such a region to be extended for 
$-1<\epsilon<0$, as we can see in Fig. \ref{F}.
In Eq. (\ref{TotalE}) providing $\epsilon\leq-1$ the fluctuations sector becomes 
unstable in analogy with the unstable region in Fig. \ref{F}.
However the configuration is also not stable providing $\epsilon > 1$, associated
with the presence of the fluctuations, even if the energy density of the 
mean-field ground state is stable.
Although the configuration is not stable for $\epsilon >1$, and $\epsilon <-1$, 
we can do the following analysis.
In the first instance we will analyze the mean field collapse region of the 
Bose-Bose mixture, i.e. when the interspecies attraction $g^{(0)}_{12}<0$ becomes
stronger than intraspecies repulsion $g^{(0)}>0$, such that $\epsilon<-1$. It is 
clear that if  $\epsilon \rightarrow -1^-$,  the second line in the energy 
density given by Eq. (\ref{TotalE}) becomes imaginary,
and it could be induce instability \cite{Petrov1}. However if $\bar{n}$ is not 
too large such hat this contribution is small enough such that its effect is only
dissipative, as a physically reasonable assumption we can neglect it
\cite{Petrov1,Rabi}.
Under the above mentioned consideration, the system exhibits a liquidlike phase
where Rabi-coupled Bose-Bose droplets with nonuniversal effects to the
interactions emerge.
The respective scaled energy density for this phase is given by
\begin{eqnarray}
\widetilde{\mathcal{E}}_{D} &=& \pi(1-|\epsilon|)\widetilde{n}^2 -
\widetilde{\omega}_R \widetilde{n}
+ \frac{32\sqrt{2\pi}}{15} \frac{[(1+|\epsilon|)\widetilde{n}]^{5/2} }
{(1+h^-_\text{fr}\widetilde{n})^2}
\nonumber \\
&\times& \Big\{1+ \frac{\widetilde{\omega}_R}
{8\pi(1+|\epsilon|) \widetilde{n}} \frac{1} {(1+h^-_\text{fr}\widetilde{n})}
\Big[2\Big(\frac{1+|\epsilon|}{1-|\epsilon|}\Big)(5+h^-_\text{fr}
\widetilde{n})+5\Big] \Big\}.
\label{Energy-drop}
\end{eqnarray}
In a attempt for fitting theoretical predictions and numerical results for 
conventional droplets  \cite{Montecarlo-Bose-Bose1}, it is employed the 
approximation $|\epsilon| = 1$ only on the fluctuations sector.
However, if we use $|\epsilon|=1$ only on fluctuations of 
Eq. (\ref{Energy-drop}), the last term becomes divergent and we have not a 
stable ground-state.
On the other hand, for $\epsilon \gtrsim 1$, and neglecting the imaginary 
contribution of Eq. (\ref{TotalE}), the resulting total real energy density is
stable. However in this scenario the effect of repulsive Gaussian fluctuations is
imperceptible regarding leading repulsive mean-field term. 
Therefore these are not relevant and can be neglected.
\begin{figure}[t]
\begin{center}
\includegraphics[height=7.5cm, width=8.5cm, clip]{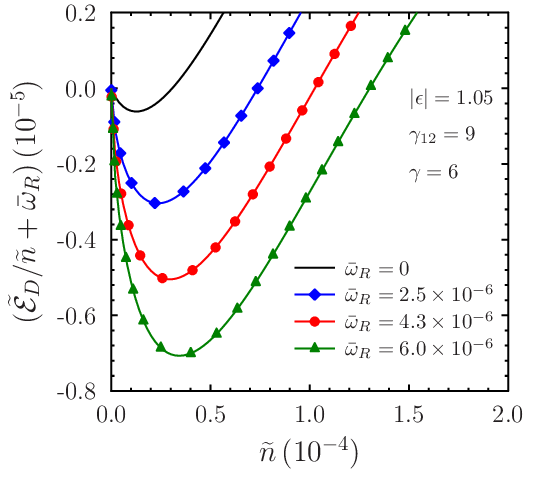}
\end{center}
\vskip -.4cm
\caption{Shifted energy per particle
$\widetilde{\mathcal{E}}_D/\widetilde{n} + {\omega}_R$  for droplets in 
Rabi-coupled Bose-Bose gases with nonuniversal corrections to the interactions 
as function of the scaled density, Eq. (\ref{Energy-drop}).
We consider the ratio $\epsilon \lesssim -1$ such that $|\epsilon|=1.05$ 
between attractive inter-species and repulsive intra-species. We set
$\gamma_{12}=9$ and $\gamma=6$.
We consider $\bar{\omega}_R = 0$ (solid black line), 
$\bar{\omega}_R = 2.5\times 10^{-6}$ (blue diamonds),
$\bar{\omega}_R = 4.3\times 10^{-6}$ (red circles), and
$\bar{\omega}_R = 6.0\times 10^{-6}$ (green triangles).
}
\label{F2}
\end{figure}

In Fig. \ref{F2} we plot the shifted energy per particle
$\widetilde{\mathcal{E}}_D/\widetilde{n} + {\omega}_R$ obtained from 
Eq. (\ref{Energy-drop}). 
We set $|\epsilon| = 1.05$, $\gamma_{12} = 6$, and $\gamma = 9$ with the 
scaled Rabi frequencies $\bar{\omega}_R = 0, 2.5\times 10^{-6},
4.3\times 10^{-6}$, and $6.0\times 10^{-6}$.
Taking into account ultracold $^{39}$K atoms these Rabi frequencies correspond to
$\bar{\omega}_R \sim 0, 40.6,69.8$, and $97.4\,$Hz, respectively.
The behavior described above is also obtained for frequencies up to
$\bar{\omega}_R=22\times 10^{-5}$ ($\sim 3572.87$ Hz for atoms of $^{39}$K).
Here we have considered the couplings in Eqs. (\ref{fr-1}), and (\ref{fr-2}).
This can be understood since taking into account only the mean-field energy per
particle the attraction causes the condensate to collapse, therefore the strong 
attraction lead us to consider the nonuniversal corrections to the attractive
interactions, such that $r_{12}\gg a_{12}$, as a relevant contribution. 
However as repulsion is very close to attraction, and with the aim of obtaining a
more complete overview of the effects of the nonuniversal corrections to the 
interactions we also include the nonuniversal correction to the repulsion such 
that $r \gg a$.
Experimental values of the s-wave scattering length and the nonuniversal 
corrections to the interactions as a function of the external magnetic field in 
ultracold $^{39}$K atoms show that it is possible to satisfy the condition 
$r \gg  a$ \cite{Montecarlo-Bose-Bose2,39K}.
In addition, if we use the couplings (\ref{Coupl1}) and (\ref{Coupl2}), the 
nonuniversal effects to the interactions on the droplet energy per particle are
not perceptible.
Droplets with  $\bar{\omega}_R=0$ and  
nonuniversal corrections to the interactions were recently considered in 
Ref. \cite{DNUC}. In such a work neglecting
the nonuniversal corrections to the repulsive interactions, and using the 
nonuniversal contribution to the attractive interactions as a fitting parameter
it is obtained a good agreement between the theoretical predictions and some 
diffusion Monte Carlo calculations (DMC) \cite{Montecarlo-Bose-Bose1}.
In Fig. \ref{F2}, and considering $a\sim 100$\AA $\,$ we have densities of order
$n \sim 10^{14}\text{atoms}/\text{cm}^3$. Our results also apply for densities of 
order $10^{13}\text{atoms}/\text{cm}^3$, experimentally obtained for conventional
droplets \cite{Exper-droplet1,Exper-droplet2}.

The increase in the driving of the population transfer between the two atomic 
levels by means of the increase of the frequency values allows us to evidence an
increasing into effective attraction of the system, both at mean-field level and
in the Gaussian fluctuations since $|\epsilon|>1$.
This leads to have a greater number of atoms in the ground-state compared to the
no presence of the Rabi coupling.

\begin{figure}[t]
\begin{center}
\includegraphics[height=9.5cm, width=12cm, clip]{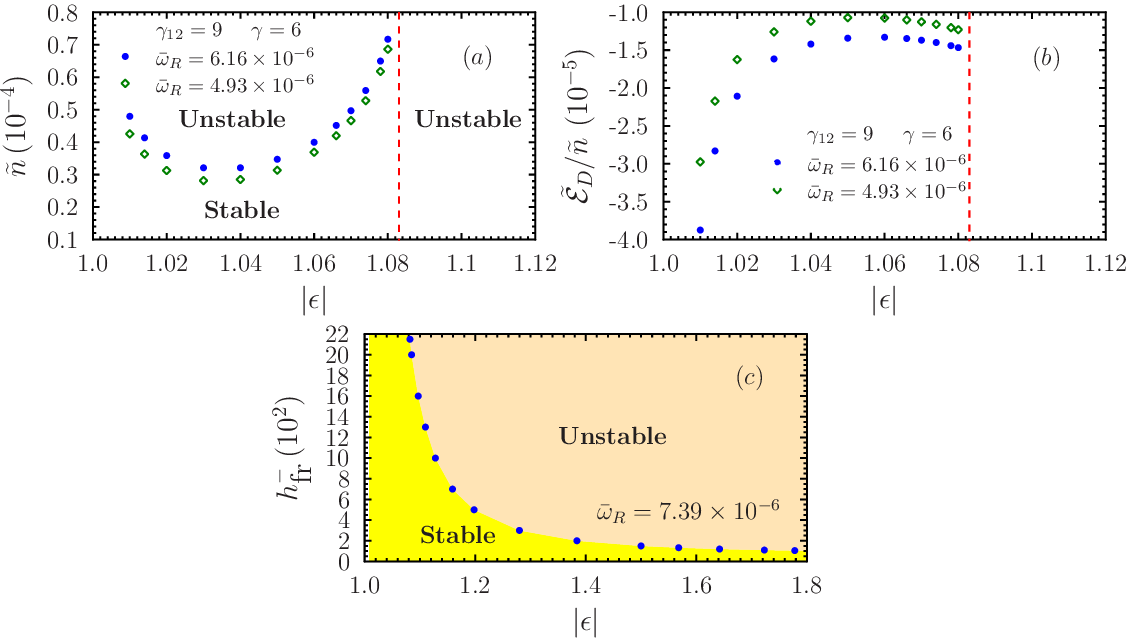}
\end{center}
\vskip -.4cm
\caption{(a)-(b)
Phase diagram for the density of the Rabi-coupled droplets with nonuniversal 
corrections to the interactions and the respective energy per particle as a 
function of the ratio $|\epsilon|$.
The nonuniversal correction to the attractions and to the repulsions are
$\gamma_{12} = 9$, and $\gamma = 6$, respectively.
We set the values of the Rabi frequency $\bar{\omega}_R$ as $6.16\times 10^{-6}$,
and $4.93\times 10^{-6}$.
The dashed red line represents the threshold value for $|\epsilon|$ as 
$|\epsilon_\text{c}|=1.083$.
(c) Phase diagram for the nonuniversal corrections to the interactions 
encoded in $h^-_{\text{fr}}>0$ as a function of $|\epsilon|$ for a frequency of
$\bar{\omega}_R = 7.39\times 10^{-6}$.
}
\label{F3}
\end{figure}

In Fig. \ref{F3} (a)-(b) we plot the phase diagram for the density of the 
Rabi-coupled droplets with nonuniversal corrections to the interactions and the 
respective energy per particle as a function of the ratio $|\epsilon|$. The 
nonuniversal corrections to the attraction and to the repulsion are fixed as
$\gamma_{12} = 9$, and $\gamma_{12} = 6$, respectively.
We set the values of the Rabi frequencies
$\bar{\omega}_R = 6.16\times 10^{-6}$ ($\sim 100$ Hz for atoms of $^{39}$K), and 
$\bar{\omega}_R = 4.93\times 10^{-6}$ ($\sim 80$ Hz for atoms of $^{39}$K).
The red line represents the critical value of $|\epsilon|$ with
$|\epsilon_\text{c}| = 1.083$, such that providing 
$|\epsilon|>|\epsilon_\text{c}|$ the droplet collapses.
In both Figs. \ref{F3} (a)-(b) we have not monotonically growing functions of the
ratio $|\epsilon|$.
For values greater than $|\epsilon|\sim 1.04$ the behavior is an expected 
consequence when the leading contribution, the attraction, increases. In other
words, we can see how increasing the atomic attraction the density of the 
droplet is increasingly and eventually this becomes unstable.
Such a behavior is also present for different values of the frequencies.
However, it is worth mentioning that due to the asymptotic behavior for 
$|\epsilon| \rightarrow 1$ related to the divergence in the fluctuation sector of 
Eq. (\ref{Energy-drop}), we also have an increasing both in density and in the
energy per particle for $|\epsilon| \lesssim 1.04$.
Similar results with the same threshold for $|\epsilon|$ have been also obtained
for frequencies up to $\bar{\omega}_R=22\times 10^{-5}$ 
($\sim 3572.87$ Hz for atoms of $^{39}$K).
In Fig. \ref{F3} (c) we include the phase diagram for the nonuniversal
corrections to the interactions encoded in $h^-_{\text{fr}}>0$ as a function of
$|\epsilon|$ for a frequency of 
$\bar{\omega}_R = 7.39\times 10^{-6}$ ($\sim 120$ Hz for atoms of $^{39}$K).
We can see how fixing both the Rabi frequency and the ratio $|\epsilon|$, and 
since $h^-_\text{fr}$ increases, the attraction exceeds of repulsion and the
effective contribution given by the nonuniversal corrections to the interactions
allows an increasing in the equilibrium density.
However, for sufficiently large values of $h^-_\text{fr}$, inelastic collisions 
decrease the number of atoms that can be accommodated in a stable ground-state 
until we no longer find a minimum in the energy density.
The system exhibits an instability and there is not a self-bound state formation, 
as it was proposed recently in droplets with nonuniversal corrections to the
interactions in Bose-Bose mixtures without Rabi coupling \cite{DNUC}.
We can also see how the threshold for the instability established by $|\epsilon|$
is smaller as the $h^-_\text{fr}$ increases, as it is expected. This can be
understood due to the fact that the leading contribution is attractive.

\begin{figure}[t]
\begin{center}
\includegraphics[height=4cm, width=12cm, clip]{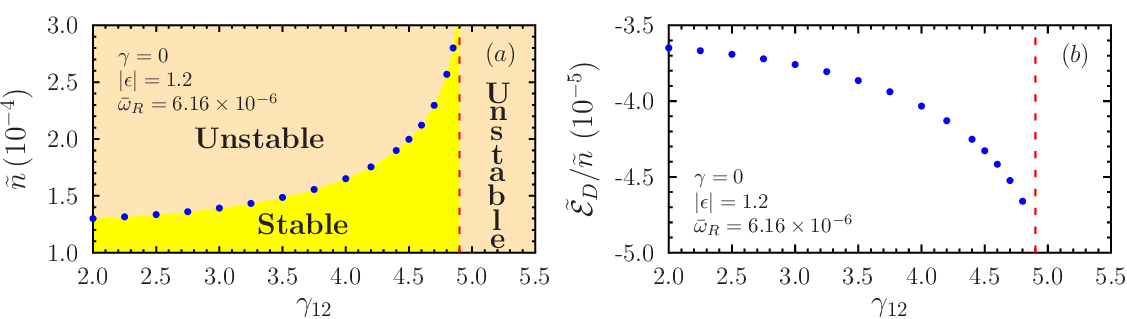}
\end{center}
\vskip -.4cm
\caption{ (a) Phase diagram for the density, and (b) energy per particle of the 
Rabi-coupled droplets with nonuniversal corrections to the interactions as 
function of the nonuniversal effects to the attractive interactions 
$\gamma_{12}$. The nonuniversal corrections to the repulsive interactions are 
fixed by $\gamma =0$. We set$|\epsilon|=1.2$  and the Rabi coupling 
$\bar{\omega}_R = 6.16\times 10^{-6}$.
The dashed red line represents the threshold value for 
$\gamma^\text{c}_{12}=4.901$.
}
\label{F4}
\end{figure}

In Fig. \ref{F4} we plot the phase diagram for the density and energy per
particle for Rabi-coupled droplets with nonuniversal corrections to the 
interactions as function of the  nonuniversal effects of the attractive 
interactions $\gamma_{12}$. 
The nonuniversal corrections to the repulsions are fixed by $\gamma =0$.
We set $|\epsilon|=1.2$ and the Rabi coupling 
$\bar{\omega}_R = 6.16\times 10^{-6}$ ($\sim 100$ Hz for atoms of $^{39}$K).
The dashed red line represents the threshold values of $\gamma_{12}$ with 
$\gamma^\text{c}_{12}=4.901$.
In other words for values greater than $\gamma^\text{c}_{12}$, the droplet 
becomes unstable by collapse.
In Fig. \ref{F4} (a)
fixing the effective attraction with constant values of $|\epsilon|$, and 
$\gamma_{12}$ and increasing the density we get an increase in the droplet
energy per particle Fig. \ref{F4} (b), until reaching a point where it is no 
longer possible to maintain a stable ground state.
The effective attraction dominates the kinetic energy, so the droplet shrinks, 
but once a threshold in $\bar{n}$ is reached the droplet becomes so attractive
that it is destroyed by inelastic collisions. As a result of it, there is 
not a minimum in energy per particle. 
This effect is also observed when the increase of the nonuniversal correction
to the attraction increases the density. The droplet is more attractive and 
eventually this becomes unstable beyond $\gamma^\text{c}_{12}$.
These results are equivalent to that presented in the Fig \ref{F3} (a)-(b),
for values not close to the asymptote $|\epsilon|\rightarrow 1$. 
This behavior has been also obtained for frequencies up to
$\bar{\omega}_R=22\times 10^{-5}$ ($\sim 3572.87$ Hz for atoms of $^{39}$K).

\section{Summary and outlook}
\label{Summary and outlook}

By considering an effective quantum field theory in the formalism of the path
integrals up Gaussia level, we derive a closed expression for the 
zero-temperature equation of state for three-dimensional ultracold Rabi-coupled 
Bose-Bose mixtures of alkali-metal atoms with nonuniversal corrections to the 
interactions in the regime of weakly Rabi frequency.
The ultraviolet divergences associated to both gapless and gapped elementary 
excitations in the Bogoliubov spectra are removed through dimensional 
regularization.
We stress and focus on the case where inter-species interactions are weakly 
attractive and subtly higher than repulsive intra-species.
Our results show that in this regime such mixtures manifest the stable formation 
of a liquidlike phase or a nonuniversal Rabi-coupled Bose-Bose droplet.
Here the nonuniversal corrections to the interactions act as an additional tool
to tune the stability properties of Bose-Bose Rabi-coupled droplets.
Through phase diagrams in which the Rabi coupling is fixed and for suitable 
values of the ratio between the inter-species scattering length and the 
intra-species scattering lengths or the nonuniversal contribution to the 
interactions we establish some conditions under which the stable droplet 
formation can be achieved.

Due to the remarkable progress in the droplet phase research, it would be 
interesting to extend the present study to one and two dimensions \cite{Rabi-E},
including dimensional crossovers \cite{Crossover}. 
In particular, in the one-dimensional model a more elaborate regularization 
process might be required for solve the ultraviolet divergence of the integrals
\cite{Rabi-E}.
Another extension of the present work could be related to the numerical study by
means of quantum Monte Carlo techniques 
\cite{Montecarlo-Bose-Bose1,Montecarlo-Bose-Bose2}.
Given the experimental relevance of trapping potentials it would be also 
interesting to consider, for example, harmonic traps to study the droplets 
obtained in the present work.
Our theoretical predictions could stimulate experimental breakthroughs taking 
into account the density values considered.
A promising candidate to the experimental observation of a nonuniversal 
Rabi-coupled droplet is the gas of $^{39}$K atoms \cite{Exp-Rabi}, by considering
the second and third lowest Zeeman states of the lowest manifold for this atoms,
i.e. 
$|\text{F}=1, \text{m}_\text{F}=-1\rangle$, and 
$|\text{F}=1, \text{m}_\text{F}=0\rangle$.
At a magnetic field $\sim 56.830(1)$G with the respective scattering lengths
$a \sim 33.4a_0$, and $a_{12}\sim -53.2a_0$, where $a_0$ is the Bohr radius.
Values of small Rabi frequency can be obtained below $6$kHz \cite{Exp-Rabi},
and some experimental values for $a,a_{12} \ll r,r_{12}$ are considered in
\cite{Montecarlo-Bose-Bose2,39K}.

\section{Acknowledgments}

I acknowledge partial support by Coordena\c{c}\~ao de Aperfei\c{c}oamento de
Pessoal de N\'ivel Superior CAPES (Brazil).
I thank Professor Sadhan Adhikari for comments on a preliminary version of the 
present work.


\begin{thebibliography}{99}

\bibitem{Artificial} J. Dalibard, F. Gerbier, G. Juzeli\=unas, A. Gostauto, and
P. Ohberg,
Colloquium: Artificial gauge potentials for neutral atoms. Rev. Mod. Phys.
{\bf 83}, 1523 (2011).

\bibitem{S1} Y. J. Lin, K. Jim\'enez-Garc\'ia, and I. B. Spielman,
Spin-orbit-coupled Bose-Einstein condensates,
Nature (London) {\bf 471}, 83 (2011).

\bibitem{S2} J. Y. Zhang, S. C. Ji, Z. Chen, L. Zhang, Z. D. Du, B. Yan,
G. S. Pan, B. Zhao, Y. J. Deng, H. Zhai, S. Chen, and J. W. Pan,
Collective dipole oscillation of a spin-orbit coupled Bose-Einstein
condensate,
Phys. Rev. Lett. {\bf 109}, 115301 (2012).

\bibitem{S3} C. Wang, C. Gao, C.M. Jian, and H. Zhai,
Spin-orbit coupled spinor Bose-Einstein condensates,
Phys. Rev. Lett. {\bf 105}, 160403 (2010).

\bibitem{S4} T.L. Ho, and S. Zhang,
Bose-Einstein condensates with spin-orbit interaction,
Phys. Rev. Lett. {\bf 107}, 150403 (2011).

\bibitem{S5} Y. Li, L. P. Pitaevskii, and S. Stringari,
Quantum tricriticality and phase transitions in spin-orbit coupled
Bose-Einstein condensates,
Phys. Rev. Lett. {\bf 108}, 225301 (2012).

\bibitem{S6} N. Goldman, G. Juzelinas, P. hberg, and I. B. Spielman,
Light-induced gauge fields for ultracold atoms,
Rep. Prog. Phys. {\bf 77}, 126401 (2014).

\bibitem{E1} E. Chiquillo,
Matter-waves in Bose-Einstein condensates with spin-orbit and Rabi couplings,
J. Phys. A: Math. Theor. {\bf 48}, 475001 (2015).

\bibitem{S7} A. Manchon, H. C. Koo, J. Nitta, S. M. Frolov, and R. A. Duine,
New perspectives for Rashba spin-orbit coupling,
Nat. Mater. {\bf 14}, 871 (2015).

\bibitem{S8} H. Zhai,
Degenerate quantum gases with spin-orbit coupling: a review
Rep. Prog. Phys. {\bf 78}, 026001 (2015).

\bibitem{E2} E. Chiquillo, 
Harmonically trapped attractive and repulsive spin-orbit and Rabi coupled
Bose-Einstein condensates,
J. Phys. A: Math. Theor. {\bf 50}, 105001 (2017).

\bibitem{E3} E. Chiquillo, 
Quasi-one-dimensional spin-orbit- and Rabi-coupled bright dipolar 
Bose-Einstein-condensate solitons,
Phys. Rev. A 97, 013614 (2018).

\bibitem{Petrov1} D. S. Petrov,
Quantum mechanical stabilization of a collapsing Bose-Bose mixture,
Phys. Rev. Lett. {\bf 115}, 155302 (2015).

\bibitem{Petrov2} D. S. Petrov, and G. E. Astrakharchik,
Ultradilute low-dimensional liquids,
Phys. Rev. Lett. {\bf 117}, 100401 (2016).

\bibitem{Rabi} A. Cappellaro, T. Macr\`\i, G. F. Bertacco, and L. Salasnich,
Equation of state and self-bound droplet in Rabi-coupled Bose mixtures,
Sci. Rep. {\bf 7}, 13358 (2017).

\bibitem{Drop-1d} G. E. Astrakharchik, and B. A. Malomed,
Dynamics of one-dimensional quantum droplets,
Phys. Rev. A {\bf 98}, 013631 (2018).

\bibitem{Swirling} Y. V. Kartashov, B. A. Malomed, L. Tarruell, and L. Torner,
Three-dimensional droplets of swirling superfluids,
Phys. Rev. A {\bf 98}, 013612 (2018).

\bibitem{Bose-Bose} A. Boudjem\^aa,
Quantum and thermal fluctuations in two-component Bose gases,
Phys. Rev. {\bf A 97}, 033627 (2018).

\bibitem{1D-3D-mixtures} E. Chiquillo,
Equation of state of the one- and three-dimensional Bose-Bose gases,
Phys. Rev. A {\bf 97}, 063605 (2018).

\bibitem{Review} Y.V. Kartashov, G.E. Astrakharchik, B.A. Malomed, and L. Torner,
Frontiers in multidimensional selftrapping of nonlinear fields and matter,
Nat. Rev. Phys. {\bf 1}, 185 (2019).

\bibitem{Rabi-E} E. Chiquillo,
Low-dimensional self-bound quantum Rabi-coupled bosonic droplets,
Phys. Rev. A {\bf 99}, 051601(R) (2019).

\bibitem{Adhikari2} S. Gautam, and S. K. Adhikari,
Self-trapped quantum balls in binary Bose-Einstein
condensates,
J. Phys. B: At. Mol. Opt. Phys. {\bf 52}, 055302 (2019).

\bibitem{Collective-excit} M. Tylutki, G.E. Astrakharchik, B.A. Malomed, and 
D.S. Petrov,
Collective excitations of a one-dimensional quantum droplet,
Phys. Rev. A {\bf 101}, 051601(R) (2020).

\bibitem{Pairing1} H. Hu, and X.-J. Liu,
Consistent theory of self-bound quantum droplets with bosonic pairing,
Phys. Rev. Lett. {\bf 125}, 195302 (2020).

\bibitem{Pairing2} H. Hu, J. Wang, and X.-J. Liu, Microscopic pairing theory of 
a binary Bose mixture with interspecies attractions: Bosonic BEC-BCS crossover
and ultradilute low-dimensional quantum droplets,
Phys. Rev. A {\bf 102}, 043301 (2020).

\bibitem{Phonon1} Q. Gu, and L. Yin,
Phonon stability and sound velocity of quantum droplets in a boson mixture,
Phys. Rev. B {\bf 102}, 220503(R) (2020).

\bibitem{Normal-superfluid} L. He, P. Gao, and Z.-Q. Yu,
Normal-superfluid phase separation in spin-half bosons at finite temperature,
Phys. Rev. Lett. {\bf 125}, 055301 (2020).

\bibitem{Boudjem} A. Boudjem\^aa,
Many-body and temperature effects in two-dimensional quantum droplets 
in Bose-Bose mixtures,
Sci. Rep. {\bf 11}, 1 (2021).

\bibitem{Bose-Bose2} N. Guebli, and A. Boudjem\^aa,
Quantum self-bound droplets in Bose-Bose mixtures: Effects of higher-order 
quantum and thermal fluctuations,
Phys. Rev. A {\bf 104}, 023310 (2021).

\bibitem{Evaporation} G. De Rosi, G.E. Astrakharchik, and P. Massignan,
Thermal instability, evaporation and thermodynamics of onedimensional liquids in
weakly-interacting Bose-Bose mixtures,
Phys. Rev. A {\bf 103}, 043316 (2021).

\bibitem{Phonon2} Y. Xiong, and L. Yin,
Effective single-mode model of a binary boson mixture in the quantum droplet 
region,
Phys. Rev. A {\bf 105}, 053305 (2022).

\bibitem{Spinor1} T. A. Yo\u{g}urt, A.  Kele\c{s}, and M.\"O. Oktel,
Spinor boson droplets stabilized by spin fluctuations,
Phys. Rev. A {\bf 105}, 043309 (2022).

\bibitem{Spinor2} T. A. Yo\u{g}urt, A.  Kele\c{s}, and M.\"O. Oktel,
Polarized Rabi-coupled and spinor boson droplets,
Phys. Rev. A {\bf 107}, 023322 (2023).

\bibitem{Criticality} L. He, H. Li, W. Yi, and Z.-Q. Yu,
Quantum Criticality of Liquid-Gas Transition in a Binary Bose Mixture,
Phys. Rev. Lett. {\bf 130}, 193001 (2023).

\bibitem{DNUC} E. Chiquillo,
Bose-Bose gases with nonuniversal corrections to the interactions: a droplet
phase,
Ann. Phys. {\bf 475} 169955 (2025). 

\bibitem{Vortices2d} 
Y. Li, Z. Chen, Z. Luo, C. Huang, H. Tan, W. Pang, and B. A. Malomed, 
Two-dimensional vortex quantum droplets,
Phys. Rev. A {\bf 98}, 063602 (2018).

\bibitem{Symm-break-1d}
B. Liu, H.-F. Zhang, R.-X Zhong, X.-L. Zhang, X.-Z. Qin, C. Huang, Y.-Y. Li, 
and B. A. Malomed,
Symmetry breaking of quantum droplets in a dual-core trap,
Phys. Rev. A {\bf 99}, 053602 (2019).

\bibitem{Exper-droplet1} C. R. Cabrera, L. Tanzi, J. Sanz, B. Naylor,
P. Thomas, P. Cheiney, and L. Tarruell,
Quantum liquid droplets in a mixture of Bose-Einstein condensates,
Science {\bf 359}, 301 (2018).

\bibitem{Exper-droplet2} G. Semeghini, G. Ferioli, L. Masi, C. Mazzinghi,
L. Wolswijk, F. Minardi, M. Modugno, G. Modugno, M. Inguscio, and M. Fattori,
Self-bound quantum droplets of atomic mixtures in free space,
Phys. Rev. Lett. {\bf 120}, 235301 (2018).

\bibitem{Exper-droplet3} P. Cheiney, C. R. Cabrera, J. Sanz, B. Naylor, L. Tanzi,
and L. Tarruell,
Bright soliton to quantum droplet transition in a mixture of Bose-Einstein
condensates,
Phys. Rev. Lett. {\bf 120}, 135301 (2018).

\bibitem{Exper-droplet4} G. Ferioli, G. Semeghini, L. Masi, G. Giusti, G.
Modugno, M. Inguscio, A. Gallem\'i, A. Recati, and M. Fattori,
Collisions of Self-Bound Quantum Droplets,
Phys.Rev. Lett. {\bf 122}, 090401 (2019).

\bibitem{Exper-droplet5} C. D'Errico, A. Burchianti, M. Prevedelli, L. Salasnich,
F. Ancilotto, M. Modugno, F. Minardi, and C. Fort,
Observation of quantum droplets in a heteronuclear bosonic mixture,
Phys. Rev. Research {\bf 1}, 033155 (2019).

\bibitem{Exper-droplet6} G. Ferioli, G. Semeghini, S. Terradas-Brians\'o, 
L. Masi, M. Fattori, and M. Modugno,
Dynamical formation of quantum droplets in a $^{39}$K mixture,
Phys. Rev. Research {\bf 2}, 013269 (2020).

\bibitem{Exper-droplet7} T.G. Skov, M.G. Skou, N.B. J{\o}rgensen, and J.J. Arlt,
Observation of a Lee-Huang-Yang Fluid, 
Phys. Rev. Lett. {\bf 126}, 230404 (2021).

\bibitem{Exper-droplet8} Z. Guo, F. Jia, L. Li, Y. Ma, J. M. Hutson, X. Cui, and
D. Wang,
Lee-Huang-Yang effects in the ultracold mixture of $^{23}$Na and $^{87}$Rb
with attractive interspecies interactions,
Phys. Rev. Research {\bf 3} 033247 (2021).

\bibitem{trapped-dip-LHY-1} F. W\"achtler, and L. Santos,
Quantum filaments in dipolar Bose-Einstein condensates,
Phys. Rev. A {\bf 93}, 061603(R) (2016).

\bibitem{trapped-dip-LHY-2} F. W\"achtler, and L. Santos,
Ground-state properties and elementary excitations of quantum droplets in
dipolar Bose-Einstein condensates,
Phys. Rev. A {\bf 94}, 043618 (2016).

\bibitem{d-LHY-1} D. Baillie, R.M. Wilson, R.N. Bisset, and P.B. Blakie,
Self-bound dipolar droplet: a localized matter wave in free space,
Phys. Rev. A {\bf 94}, 021602(R) (2016).

\bibitem{d-LHY-2} R. N. Bisset, R. M. Wilson, D. Baillie, and  P. B. Blakie,
Ground-state phase diagram of a dipolar condensate with quantum fluctuations,
Phys. Rev. A {\bf 94}, 033619 (2016).

\bibitem{Montecarlo-dip1} H. Saito,
Path-integral Monte Carlo study on a droplet of a dipolar Bose-Einstein
condensate stabilized by quantum fluctuation,
J. Phys. Soc. Jpn. {\bf 85}, 053001 (2016).

\bibitem{d-LHY-3} Abdel\^aali Boudjem\^aa,
Fluctuations and quantum self-bound droplets in a dipolar Bose-Bose mixture,
Phys. Rev. A {\bf 98}, 033612 (2018).

\bibitem{Rosensweig} H. Kadau, M. Schmitt, M. Wenzel, C. Wink, T. Maier, I.
Ferrier-Barbut, and T. Pfau,
Observing the Rosensweig instability of a quantum ferrofluid,
Nature (London) {\bf 530}, 194 (2016).

\bibitem{Self-Drop-Dy} M. Schmitt, M. Wenzel, F. B\"ottcher, I. Ferrier-Barbut,
and T. Pfau,
Self-bound droplets of a dilute magnetic quantum liquid,
Nature (London) {\bf 539}, 259 (2016).

\bibitem{Trap-Drop-Dy-1} I. Ferrier-Barbut, H. Kadau, M. Schmitt, M. Wenzel, and
T. Pfau,
Observation of quantum droplets in a strongly dipolar Bose gas,
Phys. Rev. Lett. {\bf 116}, 215301 (2016).

\bibitem{Trap-Drop-Dy-2} I. Ferrier-Barbut, M. Schmitt, M. Wenzel, H. Kadau, and
T. Pfau,
Liquid quantum droplets of ultracold magnetic atoms,
J. Phys. B: At. Mol. Opt. Phys. {\bf 49} 214004 (2016).

\bibitem{Erbium} L. Chomaz, S. Baier, D. Petter, M. J. Mark, F. W\"achtler,
L. Santos, and F. Ferlaino,
Quantum-fluctuation-driven crossover from a dilute Bose-Einstein condensate to a
macro-droplet in a dipolar quantum fluid,
Phys. Rev. X {\bf 6}, 041039 (2016).

\bibitem{Vortices3d-dip} G. Li, Z. Zhao, X. Jiang, Z. Chen, B. Liu, Boris A. Malomed, and Y. Li,
Strongly Anisotropic Vortices in Dipolar Quantum Droplet,
Phys. Rev. Lett. {\bf 133}, 053804 (2024).

\bibitem{Numeric-drop} C. Staudinger, F. Mazzanti, and R.E. Zillich,
Self-bound Bose mixtures,
Phys. Rev. A {\bf 98}, 023633 (2018).

\bibitem{Montecarlo-Bose-Bose1} V. Cikojevi\'{c}, L. V. Marki\'{c},
G. E. Astrakharchik, and J. Boronat,
Universality in ultradilute liquid Bose-Bose mixtures
Phys. Rev. A {\bf 99}, 023618 (2019).

\bibitem{Montecarlo-Bose-Bose2} V. Cikojevi\'{c}, L. V. Marki\'{c}, and J.
Boronat,
Finite-range effects in ultradilute quantum drops, 
New J. Phys. {\bf 22}, 053045 (2020).

\bibitem{Beyond LHY} M. Ota, and G. E. Astrakharchik,
Beyond Lee-Huang-Yang description of self-bound Bose mixtures,
SciPost Phys. {\bf 9}, 020 (2020).

\bibitem{Andersen} J.O. Andersen, Rev. Mod. Phys. {\bf 76}, 599 (2004).

\bibitem{Exp-Rabi} L. Lavoine, A. Hammond, A. Recati, D. Petrov, and T. Bourdel, 
Beyond-mean-field effects in Rabi-coupled two-component Bose-Einstein condensate,
Phys. Rev. Lett. {\bf 127}, 203402 (2021).

\bibitem{Nonuniversal1} E. Braaten, H.-W. Hammer, and S. Hermans,
Nonuniversal Effects in the Homogeneous Bose Gas,
Phys. Rev. A {\bf 63}, 063609 (2001).

\bibitem{Nonuniversal2} H. Fu, Y. Wang, and B. Gao,
Beyond the Fermi pseudopotential: A modified Gross-Pitaevskii equation,
Phys. Rev. A {\bf 67}, 053612 (2003).

\bibitem{Magnetic-solit-Rabi} C. Qu, M. Tylutki, S. Stringari, and Lev P.
Pitaevskii,
Magnetic solitons in Rabi-coupled Bose-Einstein condensates,
Phys. Rev. A {\bf 95}, 033614 (2017).

\bibitem{Le Bellac} M. Le Bellac, \textit{Thermal Field Theory},
(Cambridge University Press, Cambridge, 1996).

\bibitem{Stoof} H.T.C Stoof, K.B. Gubbels, and D.B.M. Dickerscheid,
\textit{Ultracold Quantum Fields} (Springer, Berlin, 2009).

\bibitem{LHY-1} T. D. Lee and C. N. Yang,
Quantum-mechanical many-body problem with hard-sphere interaction,
Phys. Rev. {\bf 105}, 1119 (1957).

\bibitem{LHY-2} T. D. Lee, K. Huang, and C. N. Yang,
Eigenvalues and eigenfunctions of a Bose system of hard spheres and its 
low-temperature properties,
Phys. Rev. {\bf 106}, 1135 (1957).

\bibitem{Larsen} D. M Larsen,
Binary mixtures of dilute bose gases with repulsive interactions at low 
temperature,
Ann. Phys. (Berlin) {\bf 24}, 89 (1963).

\bibitem{Pit-Strin} L. P. Pitaevskii and S. Stringari, 
\textit{Bose-Einstein Condensation and Superfluidity} (Clarendon, Oxford, UK, 
2016).

\bibitem{Peth-Smith} C. J. Pethick and H. Smith, \textit{Bose-Einstein
Condensate in Dilute Gases}, (Cambridge, University Press, UK, 2008).

\bibitem{Scattering1} J. Sakurai,
Modern Quantum Mechanics (Addison-Wesley Publishing
Company, Boston, 1994)

\bibitem{Scattering2} L. S. Rodberg, and M. Thaler, Introduction to
the Quantum Theory of Scattering (Academic Press, 1970).

\bibitem{On-shell} F. Lorenzi, A. Bardin, and L. Salasnich,
On-shell approximation for the s-wave scattering theory,
Phys. Rev. A {\bf 107}, 033325 (2023).

\bibitem{Adhikari} S. K. Adhikari,
Improved effective-range expansions for small and large values of scattering 
length,
Eur. J. Phys. {\bf 39}, 055403 (2018).

\bibitem{Son} D. T. Son, and M. A. Stephanov,
Domain walls of relative phase in two-component Bose-Einstein condensates,
Phys. Rev. A {\bf 65}, 063621 (2002).

\bibitem{Equal-chem-pot} C. P. Search, A. G. Rojo, and P. R. Berman,
Ground state and quasiparticle spectrum of a two-component Bose-Einstein
condensate,
Phys. Rev. A {\bf 64}, 013615 (2001).

\bibitem{Rabi1} M. Abad, and A. Recati,
A study of coherently coupled two-component Bose-Einstein condensates,
Eur. Phys. J. D {\bf  67}, 40053, (2013).

\bibitem{Modes} J. Armaitis, H. T. C. Stoof, and R. A. Duine,
Hydrodynamic modes of partially condensed Bose mixtures,
Phys. Rev. A {\bf 91}, 043641 (2015).

\bibitem{Regularization} L. Salasnich, and F. Toigo,
Zero-point energy of ultracold atoms,
Phys. Rep. {\bf 640}, 1 (2016).

\bibitem{Hooft} G. 't Hoof, and M. Veltman,
Regularization and renormalization of gauge fields,
Nucl. Phys. B {\bf 44}, 189 (1972).

\bibitem{dim-reg} H. Kleinert, and V. Schulte-Frohlinde, \textit{Critical
Properties of $\phi^4$-Theories}, (World Scientific, Singapore, 2001).

\bibitem{39K} S. Roy, M. Landini, A. Trenkwalder, G. Semeghini, G. Spagnolli, 
A. Simoni, M. Fattori, M. Inguscio, and Giovanni Modugno,
Test of the Universality of the Three-Body Efimov Parameter at Narrow Feshbach
Resonances,
Phys. Rev. Lett. {\bf 111}, 053202 (2013).

\bibitem{Crossover} P. Zin, M. Pylak, T. Wasak, M. Gajda, and Z. Idziaszek,
Quantum Bose-Bose droplets at a dimensional crossover,
Phys. Rev. A {\bf 98}, 051603(R) (2018).

\end{thebibliography}
\end{document}